# A Graphics Function Standard Specification Validator

Steven D. Fraser and Peter P. Silvester
Computational Analysis and Design Laboratory, Department of Electrical Engineering,
McGill University, Montreal, CANADA H3A 2A7

**Abstract**

A validation methodology is proposed and implemented for natural language software specifications of standard graphics functions. Checks are made for consistency, completeness, and lack of ambiguity in data element and function descriptions. Functions and data elements are maintained in a relational database representation. The appropriate checks are performed by sequences of database operations. The relational database manager INGRES was used to support a prototype implementation of the proposed technique.

The methodology supports the development of a scenario-based prototype from the information available in the specification. This permits various function sequences to be checked without implementation of the environment specified. The application of a prototype implementation of the proposed methodology to the specification of the Graphics Kernel System (GKS) software package demonstrates the practicability of the method. Several inconsistencies in GKS related to the definition of data elements have been identified.

**Keywords:** Prototype, scenario, software package, standard, validation.

## 1. Introduction

Problem definition presents a problem in itself. The question *"Has the problem been posed properly?"* is quite different from the question *"Has the problem been answered properly?"* The answer to the first question is termed validation, the answer to the second one – verification.

The validation problem considered is to determine whether the specification of a graphics standard meets certain consistency and completeness requirements both with respect to internal definitions and with regard to user requirements. This is not a simple task because the specifications (problem descriptions) to be validated are invariably written in a natural language such as English. Formal natural language processing is not directly addressed by this research. It is circumvented by means of a classification methodology.

While it is difficult to apply formal techniques, application of a systematic methodology can produce worthwhile results that will help ensure that the system specified will satisfy client requirements. There are three principles that form the basis of the methodology described: Set Theory (set membership): Predicate Calculus (*"pre"* and *"post"* conditions): and the general principles of Software Engineering. The determination of set membership is an important tool in ensuring that the static definition of the specification is both consistent and complete as far as may be determined from the information available in the specification. For example, any information referenced by the specification must be defined. This definition may occur at an arbitrary level of abstraction.

*"Pre"* and *"post"* conditions support the prototyping mechanism used to determine that the specification is consistent with client requirements. The status of information before and after the performance of a function is described by the specification. While this supports a very limited range of scenarios, it does not require information additional to that present in the specification. For example, the sequencing of functions may be examined with respect to the availability and state of information to be manipulated by the specified system. This supports valuable communications between the clients and the specifiers to develop and maintain a system specification that solves the appropriate problem.

The principles of software engineering are somewhat nebulous. The chances are good that no two software engineers would agree on their precise statement. This is somewhat different from other engineering disciplines, for example, subject to discussions on notation, most electromagnetic engineers would agree that Maxwell's equations form a required basis for work in their field. Specific software engineering principles that have been employed in the proposed technique include: template and menu driven user interfaces, incremental checks, and the control of complexity.

## 2. Terminology

A brief introduction to the terminology employed is given to permit a better understanding of both the problem and the application of the proposed methodology. This is also necessary as there is no widespread agreement on the application of a standard terminology to this area of research.

A **software package** is considered as a collection of loosely coupled functions that act on a unified database (a set of subroutines rather than stand-alone programs). A **standard** is a concept that is based on community established and agreed upon procedures and practices [1]. Standards should be viewed as an "external" input for the specification process that cannot be readily modified by implementors. In the specification of standard software packages, it is assumed that a specification will lead to many different implementations that must be compatible to the degree indicated by the specification.

**Prototypes** are used to create models of the system. There are several basic types of prototypes. In general, a prototype is a functional model used to determine requirements for a proposed system at a complexity and a cost level of perhaps one-tenth that of the actual system. Such a prototype is later discarded and replaced with the operational system. Alternatively, the prototype may also be developed and refined into the final product directly [2]. A **scenario-based prototype** [3] simulates events that would be experienced by a user of the specified system. Unlike true prototypes, that perform limited useful functions, scenarios in general permit a simulation of the sequence of specified events and input/output operations without actual production of real results.

A specification is a description of "what" the system is expected to perform. Specifications may be considered at different levels of abstraction. We have proposed a hierarchical organisation of four different levels, suitable for engineering or scientific tasks:
1. A stated need for a system that will maintain and process information.
2. A description of detailed requirements, in terms of the information processed and the functionality necessary for the support of the proposed system.
3. A detailed description of the algorithms and structures required to process and represent the requirements.
4. The executable software code to implement the requirements previously stated.

The specification of software packages is often more complex than the specification of complete end-user systems, because the highest levels of abstraction are rarely available to the specifier. Therefore, this research is primarily concerned with "Level 2" specifications. Consistency and completeness checking is logically best done at the second level, where the data, functions, and restrictions are known, but not yet obscured by detailed procedural needs. Comparison of real needs to those specified is difficult with formal methods, for there is no formal way of defining user needs [4]. Naur [5] touches on this subject when he suggests that intuition has a valid role to play in software development. The ultimate test of validation is to determine whether the system specified is acceptable. One approach would be to demonstrate the finished product to end-users who would then determine whether it was satisfactory. This is usually unacceptably costly. A reasonable compromise is to produce a scenario-based prototype.

**Validation** is the activity of comparison between "user needs" and the specification. This research assumes that the specification of interest is of type "level 2" as described previously. Validation, in this sense, may be viewed both statically and dynamically. Attributes of consistency, existence, and uniqueness checks form the basis of static checks. Dynamic checks permit the user to experiment with a prototype of the system to determine the acceptability of functionality and sequencing. Validation should not be confused with verification which is the process of comparison between the fully implemented system and the specification (Does the implemented system function as specified?).

### 3. Technique

The technique developed approaches validation as both a static and a dynamic operation. Static checking is accomplished interactively as the specification to be checked is processed. The definition of "data elements" is fundamental to this process. Each concept to be manipulated by a function must be defined at some level as a data element (or group of data elements). Functions consist of parameter, effect, and restriction definitions. General outlines of the technique have been previously described in [6] and [7]. The specification structure is illustrated in Fig. 1.

It is assumed that the specification of the software package structure may be divided into two parts: information structures and associated transformations. Information is embedded in *data elements* while transformations are described by *functions*. Each data element is classified by a *data type*.

Each function must have a unique name for reference. Functions are classified by their *type*, *level* and *state*. The function type attribute classifies functions by their general nature, e.g., as control functions or output functions. The function level determines the degree of function availability for systems that may not require a full implementation. The term *function state* is used to describe when the function may be applied to a particular combination of data element values. A function application may be possible in more than one operational state.

Three descriptors, *allocation*, *definition*, and *value*, are used to describe how much is known about each data element. All three are Boolean state indicators that are used to describe "pre-" and "post-" conditions. Upon creation of a data element the state indicators are set to *false*, so that every data element is initially considered to be unallocated, undefined, and to have no value. When manipulating rough ideas, it is easy enough to introduce a new concept and leave it undetailed. Detail is generally introduced later in the definition process. As more information becomes available during a scenario generation, at least some state indicators will change to *true*.

When a data element is first accessed by a function, it must be initialised. If this was not the case, as an input to a function it will cause a *validation exception* to prevail in the validation system described here. In other words, the original assumption is declared incorrect and the specification is considered inconsistent, incomplete, or ambiguous. The result of the output operation will, at the very least, change the allocation state indicator to true to indicate that an operation akin to allocation of memory on a computer has taken place. If the data element has been defined, then the definition state indicator is set to *true*. A data element whose state indicators are true may be used as an argument of a function. If the value of a data element is actually known, then the value state indicator is set to *true* and the appropriate value is maintained in the data element. The difference between definition and valuation must be clearly distinguished. A data element may be defined, but its value may be unknown.

Static definition sets the data element state indicators (allocation, definition, and value) to *false*. *Parameter lists* are used to describe the interfaces between functions and data elements. *Parameters* constitute the set of data elements to be manipulated by a function. They constitute the sole list of information affected by a function (assuming orthogonality of data elements). Parameters are classified in accordance with the nature of the data element referenced. A parameter may be classified as a function input or a function output.

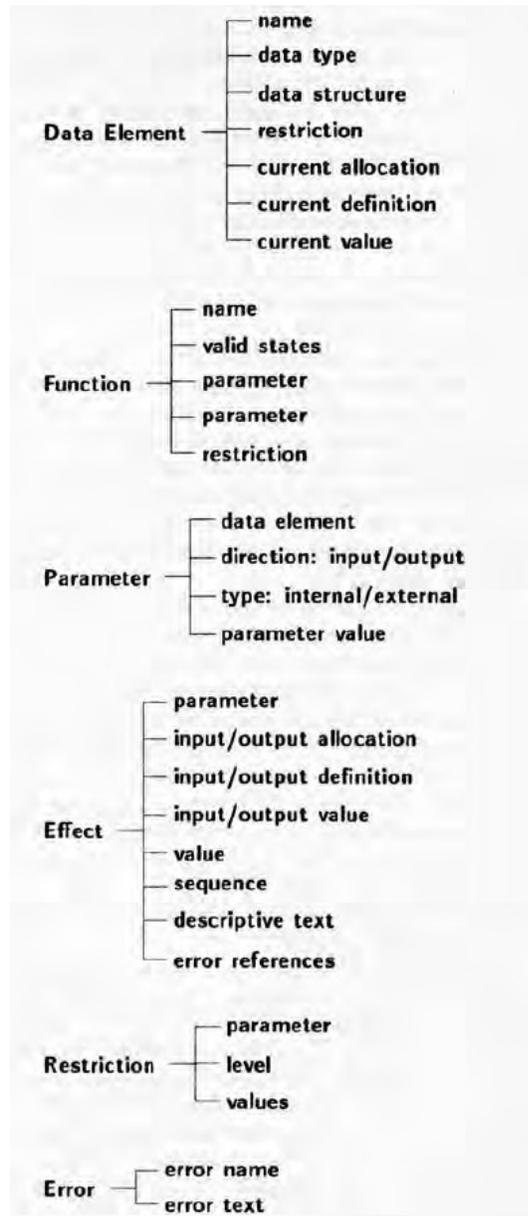

**Figure 1:** Specification Structure

It is assumed that no parameter may be classified as available for both input and output. This restriction may appear unduly strong at first glance, but it is seen on reflection to be less so, for it is the *parameter*, not the *data element*, which is restricted. While a parameter list constitutes the set of information that is passed to or from a function, other information internal to the system may also be affected. This information must be identified, even when it is internal to the system and does not appear in the external parameter list. (However, it must appear in the parameter list of some function to be considered as contained in the specification.) Thus a parameter is designated either internal or external depending on whether an explicit reference is made to it in the function parameter list.

*Effects* represent the operation that a function performs, within the limitation that no description is given of how the operation is carried out. Implementation concerns, in other words, are not considered. An effect is considered to describe the processing of a set of data elements, the function parameters. An effect places requirements on the allocation, definition, and value indicators. These indicator values form the basis of checking the dynamic interplay of the functions and data elements. Violations of the value requirements are flagged as validation exceptions during the dynamic checking process.

Effects are subdivided into five classes (by their associated text restriction): *initialise/set*: *transformation effect*: *test effect*: *test and transform effect*: and *effect not classified*. This classification is used to establish the manner in which the effect is to be dynamically simulated. It is assumed, of course, that the functions whose effects are to be described are deterministic in nature.

A *restriction* is an imposed limitation on the value that a data element or a parameter might possess. Mathematically, a restriction is a domain definition. Restrictions are imposed at two distinct levels: on data elements; and on parameters. Numeric and class membership restrictions are provided.

Numeric restrictions impose limits on the domain of validity of input. For example, a graphics device might limit coordinates to positive integers less than some specified value n, or in the range of values (1..n). However, a parameter referencing the same data element might require that input be in the domain (1..m) where m is contained in the set (1..n). Membership concerns arise in connection with enumerated data types. An enumerated data element must have one of a set of possible values, not necessarily ordered.

Dynamic specification checks are understood to operate at several levels of complexity. It is important to identify and resolve any possible conflicts between functions. Similarly, the end user must be given an opportunity to experience the system prior to its implementation. On the other hand, it is not the objective of the validation scenario process to produce a working prototype with all the features and characteristics of the specified system.

Dynamic checks and simulations are done by manipulating the definitions of functions as given in terms of the statically defined effects. The following summarises the desirable checks:
   a) Verify that the function is callable in the current operating state of the system. (It is assumed that operating state will be set in the same fashion as in the implemented system, i.e., by function calls. Similarly checks may be made dependent on the level of implementation).
   b) After presentation of the parameters to the user and input acquisition, effect processing begins. Determine that each data element described is in its correct input state, i.e., that the usage of the Boolean state indicators (allocation, definition and value) is consistent. Inconsistencies recognised in the specification (for example, the specification may indicate that a special output in the form of an error message should be issued) and those determined by checking the specification are distinguished.
   c) Natural language text associated with each effect is displayed as processed. The Boolean state indicators associated with data elements are set as prescribed.

## 4. A Case Study: GKS

A sample function specification from the Graphics Kernel System (GKS) [8] is given in Table 1. The characteristics of a function (such as that in the example) include: function name, valid states, parameters, effects, references, and errors.

> **Function Name:** INQUIRE WORKSTATION STATE
> **Valid States:** GKCL, GKOP, WSOP, WSAC, SGOP
> **Parameters:** IN workstation identifier (Name)
> OUT error indicator (Integer)
> OUT workstation state (Enumerated)
> **Effect:**
>> If the inquired information is available, the error indicator is returned as 0 and the values are returned in the output parameters.
>>
>> If the inquired information is not available, the values returned in the output parameters are implementation dependent and the error indicator is set to one of the following error numbers to indicate the reason for non-availability:
>> 7   *GKS not in proper state: GKS shall be in one of the states WSOP WSAC SGOP*
>> 20  *Specified workstation identifier is invalid*
>> 25  *Specified workstation is not open*
>> 33  *Specified workstation is of category MI*
>> 35  *Specified workstation is of category INPUT*
>
> **References:** 4.52 4.11.2
> **Errors:** none

> **Table 1:** Sample GKS Function Specification [8]

Each function is identified by a unique name, similarly parameters refer to unique data elements that exist within the software environment supported by the software package. The operating state of the function is identified by "Valid States." "Effects" are descriptions, in natural language (English), of the operations that are to be performed by the function. "References" are pointers to additional information that is provided on the software environment supported by the package. The current research maintains this as descriptive associated text. Lastly, "Errors" are text descriptions of some condition that is abnormal. "Values" are used to maintain initial settings. An exception to this is the manipulation of enumerated descriptions, such as state values. The specification of GKS [8] is a text description covering 266 pages, of this, 116 pages are devoted to function specifications (similar to that given in Table 1), 14 pages are lists of the associated data elements, and 20 pages cover function listings (listed in five different orderings). A further three pages are devoted to an "Error List." While some of these lists are not part of the GKS standard, they do provide important additional information. The important checks required to support the manipulation of a software package include:
1) the determination of the uniqueness of names referenced (functions, data elements, errors, states, levels).
2) the determination of the consistency of description.
3) the determination of the existence of references.
4) the ability to display specification information in a systematic fashion.
5) the ability to generate a "scenario" to present the specification and to provide a limited dynamic consistency checking feature.

What do the above checks represent? They assist in the formulation of two distinct problems: determination of a process to mechanically check for both static and dynamic inconsistencies in the definition of the specification; and to offer a "scenario" to display the functionality of the proposed system prior to implementation.

In a software package specification, such as that of GKS [8], a large amount of information is maintained. In the case of GKS, approximately 150 functions and 300 data elements are structured to form the graphics environment supported by GKS. The specification of GKS was targeted for implementation by many different commercial software developers. In turn each implementation of GKS will be used by many people in the graphics community for varied graphical tasks.

The maintenance of the specification in the form proposed enables inconsistencies in static definitions to be determined prior to the formalisation of the standard. Similarly, as changes are made, the descriptions maintained may be updated with minimal difficulty, while maintaining the original consistency of the specification. We note that the specification of the GKS Standard is not an isolated document. CORE [9] and PHIGS [10] are other examples of the specifications of "standard" graphics software packages.

## 5. Application of the Technique

The prototype implementation of the static and dynamic checker consists of over 11,000 lines of C code supported by the relational database INGRES on a Micro-VAX II system. This represents approximately a one year research/programming effort. The prototype was christened S.V.S.P. for "Specification Validation of Software Packages."

With the assistance of the UNIX compatible operating system ULTRIX, logs were kept of the information entered in the system interactively. This permitted the re-application of the static checker with minimal re-entry of data if necessary, e.g., in the event of a catastrophic database failure.

The prototype system maintained the environment that was used to statically check the specification of GKS, all data element definitions were examined. Data elements were checked prior to checks between function definitions and data definitions. Lastly, function sequences were checked.

### 5.1 Definition of Data Elements

Several problems were encountered in the definition of the GKS data elements. An inconsistency was discovered when it was realised that several "real data type restrictions" were substituted for "point data type restrictions." In such case pairs of real numbers were restricted to various coordinate types (NOC: *normalized device coordinates*, DC: *device coordinates*, WC: *world coordinates*). This is inconsistent with the limited numeric restrictions available for real data types. Inconsistencies in the specification of a software package suggest that not all implementations would be compatible.

The basis for GKS data structures such as tables, queues, and lists, is not consistently or completely described in the GKS specification [8]. While vectors, matrices, lists of values, arrays of values, and ordered pairs are described in 6.1 of the GKS Data Structures, structures such as the *list of normalization transformations ordered by viewport input priority (initially in numerical order with 0 highest)* are not formally defined.

Lists, queues, and tables were supported by the prototype S.V.S.P. only to the extent of checks as to the existence of classification, i.e., no manipulation was possible of such data type values by the prototype system implemented. The data type "ordered pair" was implemented in the same fashion, i.e., as a classification only.

## 5.2 Function Descriptions

The first function to be defined was the function "OPEN GKS." While two data elements are listed as external parameters, the GKS State List, the GKS Description Table, and the GKS Workstation Description Tables, are allocated and made available. This implies that over 150 data elements are allocated and defined. Physically this requires that parameters are defined for each data element, and that the necessary initialisation effects are described.

Bundles of data elements were used to facilitate this process. However, while there is a certain degree of parallelism in the definition of function parameters where the data elements chosen are all members of a bundle, effects are usually quite individual (usually all data elements in one bundle would be inputs or outputs, as opposed to a mix, similarly all parameters would be internal or external). Different data types require varied initialisation procedures: for example, real numbers require numeric values while enumerated strings may take on any sequence of characters. Therefore, "bundle" processing for effects was limited to providing a list of data elements with the user supplying additional details for each individual effect.

## 5.3 Missing Data Elements

In the processing of function specifications and their constituent parameters it was noted that some parameters had not been previously defined as data elements. Examples of this are the "control flag" and the "error indicator" in the functions CLEAR WORKSTATION and INQUIRE LEVEL OF GKS respectively. Typically this sort of parameter carries information that has been determined by some function test and is returned for the information of the application programmer.

This type of parameter definition does not conform to the methodology proposed here. The philosophy of S.V.S.P. requires a certain minimal level of information available for manipulation. While this reduces the flexibility of parameter definition, it ensures that consistency and completeness requirements can be uniformly checked. Otherwise, the parameter definition would be derived from the function definition. In this case, similar parameter definitions might not be consistent because no common data element definition would exist as a basis for comparison.

Within the GKS specification there is an inconsistent use of the data element names defined versus those used as parameters in function definitions. For example, "window limits," an input parameter in the function SET WINDOW, is not described, although the data element "window" is described as required information for every entry in the "list of normalization transformations ordered by viewport input priority."

The specification of a standard function software package may be used as a basis for implementation and as an authoritative source for documentation (as in the case of the specification of GKS). Therefore, it is imperative to minimise ambiguities. The rationalisation of the "window" ambiguity described above required a detailed examination of data element definitions in the GKS standard to eliminate other possibilities, e.g., other data elements that had similar names.

## 5.4 Function Sequences

Various function sequences were tested. State transition functions correctly manipulated the state data element. However, the implementation of conditional tests was limited. Conditional tests are made on actual data element values. In many cases values become meaningful only in an implemented system. The cost of rapid prototype development should not become so expensive as to approach the cost of implementation, otherwise the prototype becomes simply the first implementation. Implementation of GKS in INGRES is not a desired goal.

The dynamic simulation of GKS functions enabled a satisfactory evaluation of function callability. Initialisation functions, such as **OPEN GKS**, proved somewhat tedious in the display format used. S.V.S.P. does not support the definition of grouped functions. A grouped function is used to denote a function that is defined as a group of previously defined functions. **Emergency Close GKS** is an example of such a function. **Emergency Close GKS** consists of the following functions: **Close Segment**; **Update Workstation**; **Deactivate Workstation**; **Close Workstation**; and **Close GKS**.

In terms of response time, the performance of S.V.S.P. was not entirely satisfactory. INGRES has been used to support other prototyping systems with similar difficulty [11].

## 6. CORE and PHIGS

CORE was chosen for examination, as it was a proposed standard that preceded GKS by a number of years. The CORE System Committee was created in 1976 at the SIGGRAPH Conference in Philadelphia and by the spring of 1977 at least two implementations of the proposed standard were in progress. The CORE system does not reference a "general-purpose data structure" [12, p. 130]. This makes it quite different from the general structure of GKS, and therefore it is not possible to validate CORE in the same fashion as GKS (it is a Level "1" specification in our terminology).

The ANSI standard for PHIGS (Programmer's Hierarchical Interactive Graphics Standard) was examined briefly [10]. PHIGS was developed to satisfy requirements of application programs to permit modification of 2-D and 3-D graphical data, manipulation of geometrically related objects, and rapid dynamic articulation of graphical entities [10].

The general arrangement of PHIGS function definitions is very similar to those of GKS (it is a Level "2" specification in our terminology). The same *parameter* and *effect* natural language notation is used in PHIGS.

## 7. Summary and Observations

### 7.1 Contributions

This research has implemented, on an experimental basis, a proposed methodology to check the consistency and completeness of *standard* function specifications, an example of such being the *natural language* function specifications of GKS [8]. The validation process does not explicitly require the translation of the natural language concerned into a formal notation. Specific contributions of this research are:
1. The development of a structure to maintain descriptions of specified data elements directly derived from a natural language (English) standard specification. This structure enabled the following consistency checks to be performed:
    a) consistency between data element value restrictions and parameter restrictions;
    b) consistency in the naming of data elements and their association with function parameter and effect descriptions; and
    c) consistency with user requirements by means of various interactive displays.
2. Support of function scenario generation that checks for validity of sequences of function calls where the validity of such "calls" is determined by some predefined state structure. The scenario presented is based only on the information present in the specification.
3. A prototype system (S.V.S.P.) was developed that implemented the consistency and completeness checks proposed.
4. Inconsistencies in the specification of GKS functions, e.g., parameter/data element description inconsistencies, were identified using the prototype system S.V.S.P.

## 7.2 Observations

The method implemented is functional, however some of the operations are awkward due to the prototypical nature of the system developed. While future developments suggested above are not trivial, they follow naturally from the work described here. However, no provisions have been made for the specification of non-deterministic effects.

One of the most difficult operations performed by the validation process is to classify and process effect descriptions. It is reiterated that only a limited amount of information is available for validation purposes, i.e., a description of the specified functions and their associated data elements.

The area of application is limited because the outlined methodology is proposed as a validation technique and not as a development tool. The difference between validation and development is somewhat one of perspective: if small systems are considered, an implementor could argue that both are part of the system design and implementation. However, when *standard* specifications are examined, it would be beneficial to establish a set of acceptance criteria in terms of consistency and completeness requirements to avoid problems related to ambiguity.

The manipulation of *values* is supported to the limited extent of initialisation processes internal to function effects. The dynamic simulation process does not support manipulation of input values, i.e., actual parameter values cannot be entered. It is proposed that the manipulation of such actual values would require information in addition to that provided in the specification.

It is acknowledged that the *effect* descriptions (natural language text associated with a function parameter) manipulated are somewhat vague. However, that is the limit of the information available. Several notations were examined that would have required a translation process from a natural language to a formal notation. That would have effectively altered the given specification.

The prototype system was intended as a test bed for the specification validation structure proposed by this research. It does not support all the options required for a marketable commercial product, nor was such the objective of this research. The most often asked question about the prototype system was: "Does the system do this ..." and the answer most frequently given was: "No, but ..." The object of the prototype system was to determine the feasibility of the methodology. The authors believe that this objective was satisfactorily achieved. This validation process made it possible to determine whether satisfactory consistency, completeness, and ambiguity requirements had been achieved. The validation process proposed will continue to be of value as long as natural language specifications continue to be written and used as the basis for major software package developments. It is reiterated that "graphics" in the title was used to denote the narrow range of application examples examined. The availability of the specification of GKS [8] gave considerable direction to the type of specification examined by this research.

## 8. Summary

The maintenance of a graphics function standard specification, such as that of GKS, in the form proposed enables inconsistencies in static definitions to be determined prior to the formalisation of the standard. Similarly, as changes are made, the descriptions maintained may be updated with minimal difficulty, while maintaining the original consistency of the specification.

The proposed system maintains an environment that can be used to statically check the specification of a software package. Components of the specification are entered individually. If an inconsistency is detected, it is reported. To continue the definition process, the inconsistency must be rationalised, e.g., if a function references a non-existent data element, the data element definition must be created (the alternative is to avoid the reference). Dynamic checks are made by presenting the function components, e.g., parameters, effects, associated errors, data element/parameter restrictions, and state validity. Limited manipulation of actual values may be performed, however this manipulation is generally limited to enumerated data types. "Effects" are bound to the appropriate parameter(s) and are displayed as text statements. As each "effect" is processed, the status of bound data elements (as referenced by the parameters) is checked for consistency with the status expected by the "effect."

The current implementation of the static and dynamic checker runs to approximately 11,000 lines of C code supported by the relational database INGRES. It is somewhat restricted in its means of modifying a previously entered specification. It is anticipated that the implementation of a true "modify" feature will be a priority in future work.

The limitations of scenario-based prototyping should be recognised. In particular, the temptation must be resisted to make the scenario too good, lest it begin to approach a true prototype implemented in the wrong language.

After extensive trials, the scenario-based prototype approach described appears to be satisfactory for locating inconsistencies – and perhaps more importantly, inconveniences that are not formally erroneous – in software package specifications. To conclude, several observations may be stated:

a) Consistency checks are necessary to ensure satisfaction. They are not a futile chase of the obvious, even if the system specified is only moderately complex.
b) The activity of processing specifications can easily become tedious, particularly if system response time is a factor. This tedium is difficult to avoid, however it can be minimised when it is relatively simple to manipulate a specification.

## 9. Acknowledgements

The authors wish to acknowledge the financial support of the Natural Sciences and Engineering Research Council of Canada.

[7] FRASER, S.D. and SILVESTER, P.P.: "A software package specification validation process," in *IEE Computer-Aided Engineering Journal*, Vol. 3, No. 5, October 1986, pp. 202-206. https://doi.org/10.1049/cae.1986.0051
[8] ACM SIGGRAPH, GKS: Graphical Kernel System, *Computer Graphics*, February 1984. https://doi.org/10.1145/988572.988596
[9] ACM SIGGRAPH, Status Report of the Graphic Standards Planning Committee, *Computer Graphics*, Vol. 13, No. 3, August 1979.
[10] American National Standard for the Function Specification of the Programmer's Hierarchical Interactive Graphics Standard (PHIGS), *X3H31/8203R02 PHIGS Baseline Document*, ANSI 1983.
[11] PENEDO, MARIA H.: Prototyping a Project Master Data Base for Software Engineering Environments, in *Proceedings of the ACM SIGSOFT/SIGPLAN Software Engineering Symposium on Practical Software Development Environments*, edited by Peter Henderson, in ACM SIGPLAN Notices, Vol 22., No 1., January 1987, pp. 1-11. https://doi.org/10.1145/390012.24209
[12] ACM SIGGRAPH, Status Report of the Graphics Standards Planning Committee, *Computer Graphics*, Vol. 12, Nos. 1-2, June 1978.
Page 12 | 12